\documentclass[prl,amsmath,amssymb,floatfix]{revtex4}
\usepackage[latin9]{inputenc}
\usepackage{color}
\usepackage[caption=false]{subfig}
\usepackage{verbatim}
\usepackage{graphicx}
\usepackage{esint}
\usepackage{epstopdf}

\makeatletter
\@ifundefined{textcolor}{}
{%
 \definecolor{BLACK}{gray}{0}
 \definecolor{WHITE}{gray}{1}
 \definecolor{RED}{rgb}{1,0,0}
 \definecolor{GREEN}{rgb}{0,1,0}
 \definecolor{BLUE}{rgb}{0,0,1}
 \definecolor{CYAN}{cmyk}{1,0,0,0}
 \definecolor{MAGENTA}{cmyk}{0,1,0,0}
 \definecolor{YELLOW}{cmyk}{0,0,1,0}
}

\usepackage{epsfig}% Include figure files
\usepackage{dcolumn}% Align table columns on decimal point
\usepackage{bm}% bold math
\def\(({\left(}
\def\)){\right)}
\def\[[{\left[}
\def\]]{\right]}

\newcommand{\be}{\begin{equation}}
\newcommand{\ee}{\end{equation}}
\newcommand{\bea}{\begin{eqnarray}}
\newcommand{\eea}{\end{eqnarray}}

\makeatother

\begin{document}

\title{A model with Darwinian dynamics on a rugged landscape}

\author{Tommaso Brotto$^{1,2}$, Guy Bunin$^{3}$ and Jorge Kurchan$^{1}$}

\affiliation{
$^{1}$Laboratoire de Physique Statistique, \'Ecole Normale Sup\'erieure, PSL Research University; 
Universit\'e Paris Diderot Sorbonne Paris-Cit\'e; Sorbonne Universit\'es UPMC Univ Paris 06; CNRS; 24 rue Lhomond, 75005 Paris, France.\\
$^{2}$ Dipartimento di Fisica, Universit\`a degli Studi di Milano,
Via Celoria 16, 20133 Milano, Italy. INFN, Sezione di Milano, Via Celoria 16, 20133 Milano, Italy\\
$^{3}$ Massachusetts Institute of Technology, Department of Physics,
Cambridge, Massachusetts 02139, USA}

\begin{abstract}
We discuss the population dynamics with selection and random diffusion, keeping the total
population constant,  in a fitness landscape associated with  Constraint
Satisfaction, a paradigm for difficult optimization problems.
We obtain a phase diagram in terms of the size of the population and the diffusion rate, with a glass phase
inside which the dynamics keeps searching for better configurations, and outside which deleterious `mutations' spoil 
the performance. The phase diagram is analogous to that of dense active matter in terms of temperature and drive.
\end{abstract}
\maketitle
\vspace{0.5cm}

\section{Introduction}

\vspace{0.5cm}

Optimization problems -- finding the minima of complicated functions -- are ubiquitous in science.
Statistical Mechanics has proved to be an extremely powerful tool to analyze such problems and the associated algorithms.
This is based on the recognition that the energy function  of glassy systems are archetypical rugged landscapes, 
and  that the annealing and aging of real glasses are nature's way to minimize the energy. Simulating the annealing
procedure for artificial optimization problems is a robust and quite effective method  \cite{Kirkpatrick}.

Darwinian dynamics may be viewed as an alternative method  to optimize a function -- in this case maximizing the `fitness' -- clearly
also widespread in nature.  This has been long recognized, and the literature on artificial `Genetic Algorithms' is vast \cite{genetic}.
The principle is rather different from that of annealing: instead of the algorithm  searching actively
for a better situation (a `Lamarckian' strategy), it just produces `clones' that mutate randomly and are later selected 
according to their fitness.
 Because the connection between Darwinian dynamics and physical evolution  is less 
obvious than in the case of annealing, the implications of  physics to such problems has been 
much less studied. Although there have been statistical mechanical  models of evolution (see, for example 
\cite{pedersen_long_1981,sibani_evolution_1999,seetharaman_evolutionary_2010,saakian_eigen_2004,saakian_evolutionary_2009,saakian_solvable_2004})
using the knowledge of universal  glassy features has been much less exploited \cite{franz_evolutionary_1993}.

Evolutionary programs appear naturally in physics when one models the (imaginary-time) Schroedinger Equation,
a technique known as `Diffusion Monte Carlo' \cite{anderson}, and also in the efficient calculation of
large deviations \cite{giardina}, but they may of course also be used as an alternative to Simulated Annealing for the
minimization of any cost function.
Evolutionary  dynamics has also been studied {\em per se}: the  Quasispecies Model  \cite{eigen} being perhaps the best-known example.
In these three cases, the better understood situation is the limit of large number of individuals.
However, as we shall argue below, when the dynamics takes place in a rugged landscape,
 the consequences of the finite size of the population become important  after a short (logarithmic in the size)
time-scale. This leads us to studying a dynamics in which the number of individuals $N$ is finite,
and for simplicity is kept fixed
by randomly decimating the population: a Moran process.

The dynamics of a population of $N$ individuals reproducing and undergoing
random mutations and selection has long been recognized to bear a
resemblance with a system driven by a `fitness potential', with an element
of `noise' given by random fluctuations that are  larger, the smaller
the total population (see, e.g. \cite{crow_introduction_1970,peliti_introduction_1997}).
However, the stochastic dynamics of a system in contact with a thermal
bath satisfy the relation of `detailed-balance' -- the condition that
the bath is itself in thermal equilibrium -- obviously not applicable
in general to an evolutionary dynamics with mutation and selection.
A known exception happens when the population is dominated by a single
mutant at any time, whose identity changes in rare and rapid `sweeps'
in which a new mutant fixes \cite{sella_application_2005,mustonen_fitness_2010},
see Fig. \ref{concu1c}. It turns out that in that
special case \cite{berg_stochastic_2003,berg_adaptive_2004,sella_application_2005,barton_application_2009},
there is a correspondence that we shall exploit to understand some features of the phase diagram.

In this paper we shall study the Darwinian dynamics in an archetypical constraint optimization problem (Satisfiability: KSAT and XORSAT).
 Our purpose is not to propose
this model as a relevant  metaphor for biology (there are many references on this,
see for example \cite{pedersen_long_1981,sibani_evolution_1999,seetharaman_evolutionary_2010,saakian_eigen_2004,saakian_evolutionary_2009,saakian_solvable_2004}),
but rather to work out the details in a nontrivial case. A complete analytic
solution for the population dynamics in these models is perhaps possible,
but seems like a daunting task.

\section{The Model}
\label{XorSatExample}

We shall consider a population of individuals assumed to be
independent, their internal states being  denoted  $a=1,...,2^L$'. Each has on average $\lambda_{a}$ offspring per
unit time. 
The total  number $N$ is kept constant -- or in some cases slowly varying -- by decimating 
or `cloning' randomly chosen individuals at the necessary rate, a Moran process  \cite{moran_statistical_1962}.
 The probability
of mutation per generation a state $a$ to a state $b$ is $\mu_{ab}$,
so that mutation times are random with average $\tau_{ab}=1/\left(\lambda_{a}\mu_{ab}\right) $.
In the literature, either the probabilities $\mu_{ab}$ or the times
$\tau_{ab}$ are often taken identical for all allowed mutations. We shall adopt here  $\tau_{ab}=\tau_0 \; \forall {ab} $

The evolution is described by a time-dependent distribution of types
$\{n_{1},...n_{2^{L}}\}(t)$, with $\sum_{i}n_{a}(t)=N$. Initial
conditions need to be specified, such as a population containing a
single type, or a random selection of states for the $N$ individuals.
We represent the internal state of an individual using Boolean variables: ${\bf s}^a=\{s_{1}^a,...,s_{L}^a\}$ 
taking values $s_i^a=0,1$. 
The fitness functions we use
 are standard spin-glass benchmarks, whose landscape properties
have been extensively studied \cite{kauffman_metabolic_1969}. 
It is constructed as follows: there are
$\alpha L$  clauses $\nu$ with $K=3$ variables, of the form
$(s_{i_{1}^{\nu}}\vee\overline{s_{i_{2}^{\nu}}}\vee s_{i_{3}^{\nu}})$ where
both the $(i_{1}^{\nu},i_{2}^{\nu},i_{3}^{\nu})$ chosen for each clause
-- and the fact that the variable is negated or not -- are decided
at random once and for all. The  Random K-SAT and Random Xor-SAT
take the form, for example: 
 
\begin{center}
OUTPUT = $(s_{18}\vee\overline{s_{3}}\vee s_{43})\wedge(s_{1}\vee s_{45}\vee\overline{s_{31}})...\wedge(\overline{s_{51}}\vee s_{7}\vee\overline{s_{8}})$
\emph{~~(SAT)}
\par\end{center}

\begin{center}
~~~~~~OUTPUT = $(s_{18}\veebar\overline{s_{3}}\veebar s_{43})\wedge(s_{1}\veebar s_{45}\veebar\overline{s_{31}})...\wedge(\overline{s_{51}}\veebar s_{7}\veebar\overline{s_{8}})$
\emph{~~~(XorSAT)}
\par\end{center}

If we assume that each clause has a multiplicative effect on the reproduction
rate $\lambda$, this suggests we use an  additive form for $E$

\begin{equation}
-\ln\lambda=\frac{1}{L}\sum_{a=1}^{\alpha L}[\mbox{error in clause }\nu] \equiv \frac{1}{L}E\label{ftn}
\end{equation}

The factor $\frac{1}{L}$ sets the  scale.
 We  work in a regime with $\alpha=6$:
for such a number of clauses the system virtually never has a solution
where all clauses are satisfied, i.e. $E>0$. The landscape is rugged
and the minima are separated and extremely hard to find.

\begin{figure}
\centering \includegraphics[width=0.4\textwidth]{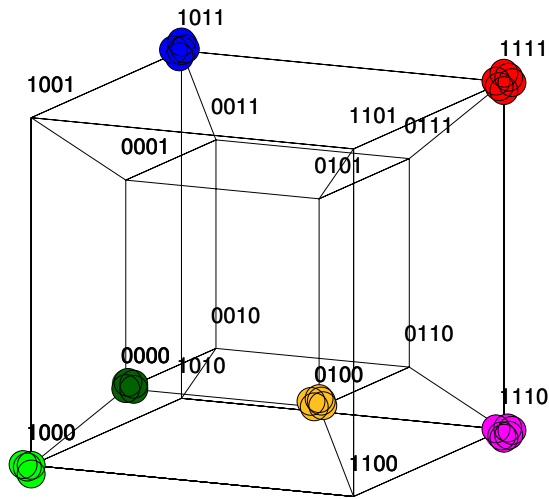} \protect\protect\protect\protect\caption{Genetic algorithm for SAT or XORSAT on the 4-dimensional hypercube. \textcolor{black}{Each vertex has a given fitness value. Individuals reproduce with the rate determined by the vertex they are in, and mutate by diffusing to connected vertices.}}
\label{cube} 
\end{figure}

The dynamics of the $N$ individuals, each identified by a vector ${\bf s}^{individual}$
is  obtained 
 by flipping randomly one of their components, in other words it is a diffusion on 
$L$ dimensional hypercube (Fig \ref{cube}), where they reproduce
or die according to the SAT or XORSAT fitness rule.

\section{A brief digression: the House of Cards Model}
\label{EmergThermal}
\vspace{0.5cm}

In order to see which are the good state parameters, and also to make this discussion less abstract, we shall first briefly 
 review a concrete example about which much is known: the `House of Cards' model \cite{kingman}. We consider again states
$a=1,...,2^{L}$ with log-fitnesses $\ln \lambda = - E/L$ distributed according to a Gaussian
distribution (a choice inspired by the Random Energy Model \cite{derrida_random-energy_1980,neher_emergence_2013},
see below).

 \begin{eqnarray}
p(E) & = & {\cal L}e^{-E^{2}/2L}\hspace{1cm}\bar{p}(\lambda)=\frac{dE}{d\lambda}\;p(E)
\end{eqnarray}
The mutation rates $\mu_{ab}$ are \textcolor{black}{identical} for all
pairs $ab$  with $\mu=\sum_{b}\mu_{ab}$,
so an individual may jump between any two states.
\begin{comment}
Note the normalization in (\ref{norma}): the factor $J$ is arbitrary,
and we shall set it to one 
\end{comment}
\begin{figure}
\centering \includegraphics[width=0.8\textwidth]{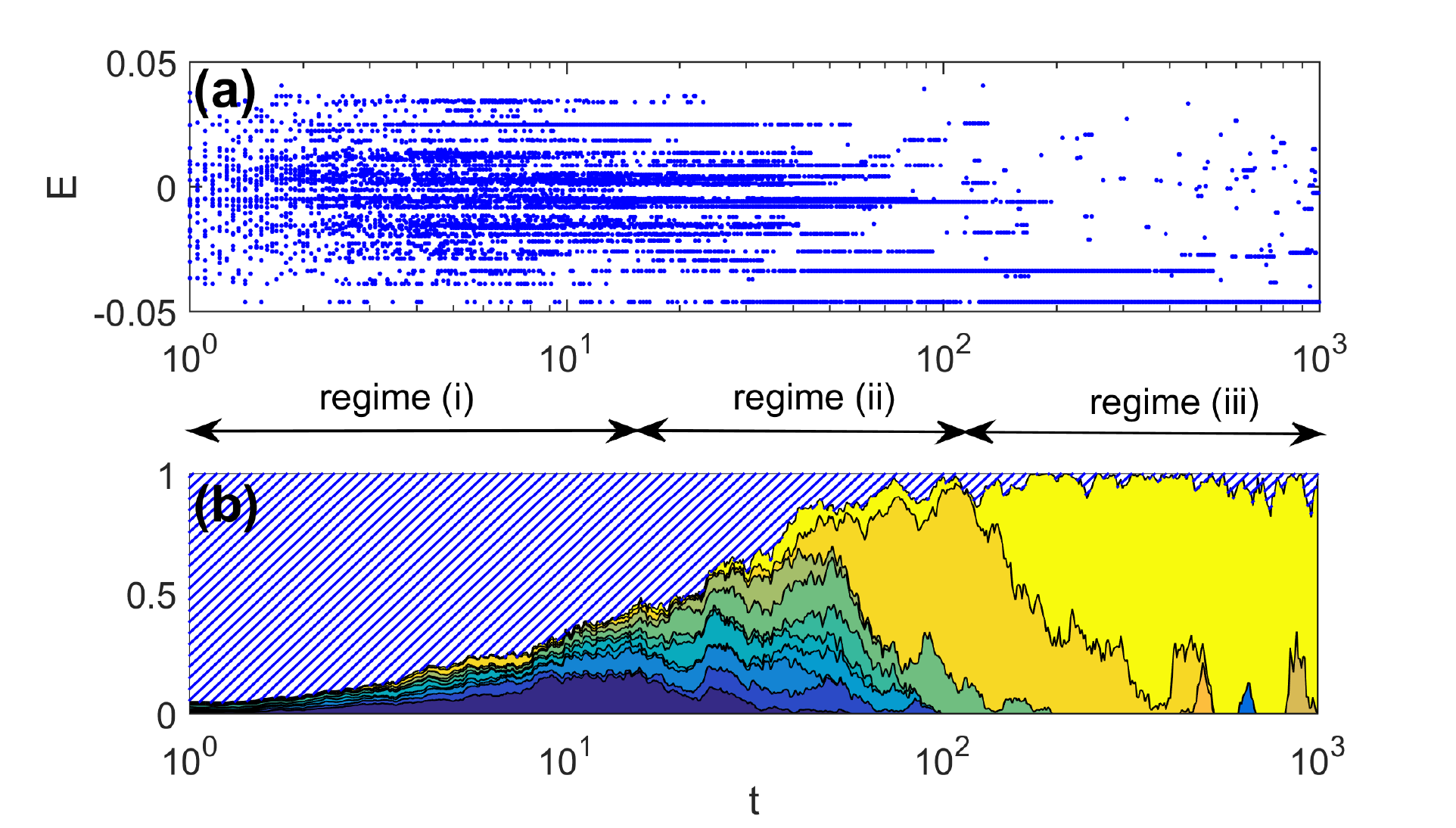}
\protect\protect\protect\protect\caption{Evolution in the REM House-of-Cards model. (a) The energy of individuals.
For clarity only 1 in 20 individuals is shown. (b) Evolution of the
largest sub-populations, as in Fig. \ref{concu1}. Model parameters  
$N=500,\,L=2500,\,\mu=1/50$.}
\label{evo} 
\end{figure}
The evolution of this system is depicted in Fig \ref{evo}, where initially each
individual is chosen randomly. The system traverses through 
several regimes: 

{\em i) Continuous population:} essentially all the population
is in states $a$ such that $1\ll n_{a}\ll N$, for all other states
$n_{a}=0$. One may treat the problem in terms of a continuous approximation
$\rho(\lambda)$ corresponding to the fraction of individuals having
$\lambda_{a}$ between contained between $\lambda$ and $\lambda+d\lambda$,
using the Replicator Equation \cite{sigmund}. We have:
\begin{equation}
\dot{\rho}(\lambda)=[\lambda\left(1-\mu\right)-\langle\lambda\rangle]\rho(\lambda)+\mu\lambda\bar{p}(\lambda)\hspace{1.5cm}with\hspace{1.5cm}\langle\lambda\rangle=\int d\lambda\,\rho(\lambda)\,\lambda\label{evolution-1}
\end{equation}
where $\mu$ \textcolor{black}{is the total probablity of mutating out of the interval $[\lambda,\lambda+d\lambda]$,}  and $\bar{p}(\lambda)$ is the
density of states.

\vspace{0.1cm}

{\em ii) Concurrent mutations regime }\cite{desai_beneficial_2007}:
Finite population size effects cannot be neglected even if the
population starts at regime (i), because they begin to show up at
times of order $\ln N$. Here a finite fraction of all individuals
are concentrated in a finite number of types  (Fig. \ref{evo})
competing for domination (strong clonal interference). (See \cite{desai_beneficial_2007,rouzine_solitary_2003,rouzine_traveling-wave_2008}
and \cite{good_distribution_2012}, especially Refs. [23-29] therein).

\vspace{0.1cm}

{ \em iii) Successional Mutation Regime}: The system settles into
a regime in which the majority of individuals belong to a single type.
Some of these individuals mutate, most often deleteriously, and die,
accounting for a constantly renewed population `cloud' of order $\mu N$
outside the dominant sub-population. Every now and then, an individual
mutates to a state that is more fit, in which case it may spread in
the population until completely taking over (fixation). There are,
in addition, events in which the entire population may get fixated to a mutation that is (slightly) less fit: these extinction events
are exponentially rare in $N$. In this regime, it is easy to compute
the probability for a new mutation to appear and fix in an interval
of time $\delta t$, large with respect to the fixation time, small
compared to the time between successive fixations \cite{crow_introduction_1970}:
\begin{equation}
P({\mbox{fixed in }}a\to{\mbox{fixed in }}b)\equiv P(a\rightarrow b)
=N\lambda_{a}\mu_{ab}\delta t\;\frac{\frac{\lambda_{a}}{\lambda_{b}}-1}{(\frac{\lambda_{a}}{\lambda_{b}})^{N}-1}
=N\lambda_{a}\mu_{ab}\delta t\; \frac{e^{\left(E_{b}-E_{i}\right)/L}-1}{e^{\left(E_{b}-E_{a}\right)N/L}-1}.\label{mc-1}
\end{equation}
where
\begin{equation}
E_{a}\equiv-L\ln\lambda_{a}\:,\label{eq:E_def}
\end{equation}
 will turn out to play a role analogous to that of an energy. The normalization is chosen  to make quantities of interest, such are
changes in fitness, of order one. (Here we have assumed that $\mu L$ is small, so we have neglected the `cloud' of deleterious mutations.)
A population will evolve in this regime whenever mutations are rare
\cite{mustonen_fitness_2010,mustonen_molecular_2008} because few useful mutations are offered in any generation \footnote{
If one allows for many mutations to exist, while still having a single
dominant population at almost all times, a somewhat different regime
is obtained \cite{desai_beneficial_2007}. Then Eq. (\ref{mc-1})
no longer holds due to the population `cloud' of deleterious mutations.
If these mutants do not reproduce ($\lambda=0$), Eq. (\ref{mc-1})
may be mended by considering an effective $N_{eff}=N-N_{cloud}$,
but for more general deleterious mutations a simple prescription is
hard to give. However, this correction is small when mutation rates
are low $\mu\ll1$ (but not necessarily very low $\mu N\ll1$). More
precisely, Eq. (\ref{mc-1}) holds when the fraction of deleterious
mutations is small, $\frac{\mu\lambda}{\lambda-\lambda_{del}}\ll1$,
where $\lambda$ is the fitness of the dominant population and $\lambda_{del}$
is a typical fitness of deleterious mutations.}.

\begin{figure}
\centering \includegraphics[angle=0,width=5cm]{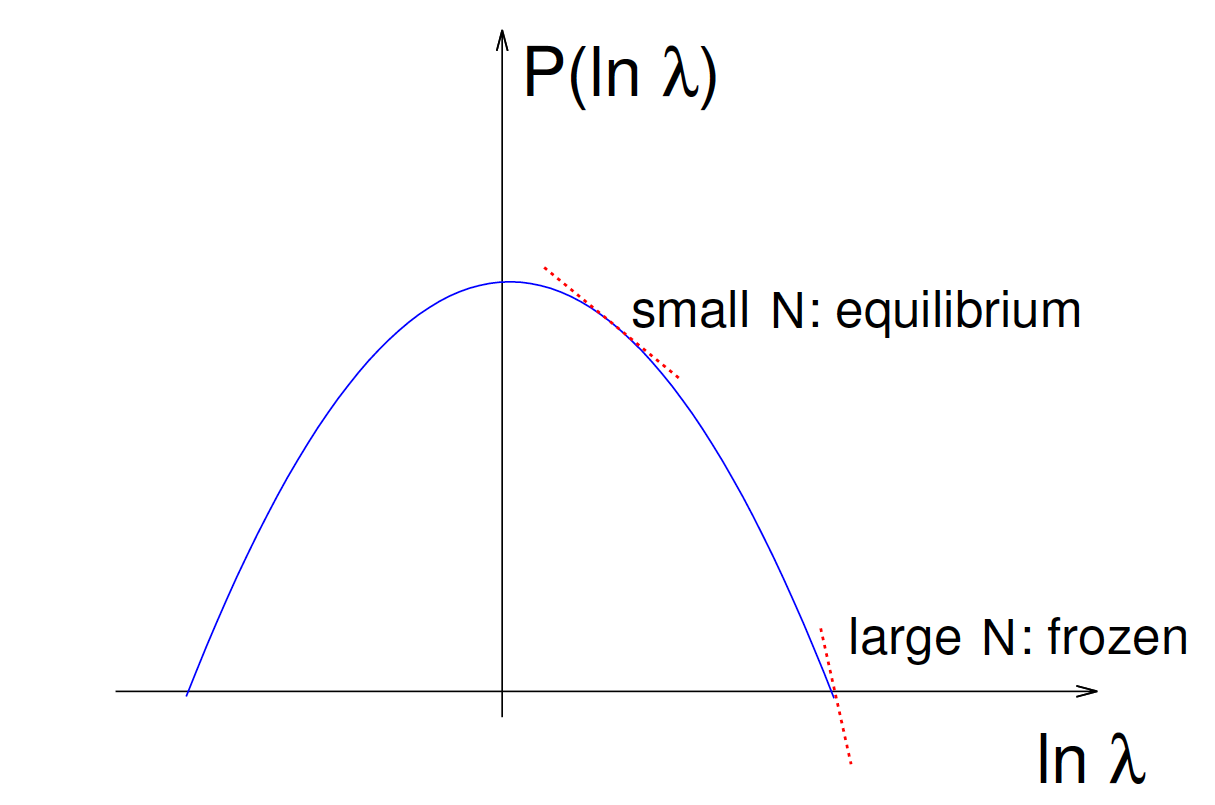} \protect\protect\protect\protect\caption{Equilibrium for the REM/House of Cards model.}
\label{remsketch} 
\end{figure}

\pagebreak

\vspace{0.5cm}

\textbf{Emergence of Detailed Balance  in the successional regime}

\vspace{0.5cm}

In the context of  the House-of-Cards model, consider now {\em for the successional
mutation regime} the `meta-dynamics' of the dominant sub-population,
considered as a single entity, neglecting the relatively short times
in which the system is not concentrated into a single type (the {{}
fixation processess}). Since in this example $\mu_{ab}=\mu_{ba}$:
\begin{equation}
\frac{P(a\to b)}{P(b\to a)}=\left(\frac{\lambda_{b}}{\lambda_{a}}\right)^{N-2}=e^{-(N-2)[\ln\lambda_{a}-\ln\lambda_{b}]}=e^{-\frac{N-2}{L}[E_{b}-E_{a}]}
\end{equation}
This corresponds to a process with detailed balance and temperature
$T=\frac{L}{N-2}$ and energies $E_{a}$ \cite{berg_adaptive_2004,berg_stochastic_2003,sella_application_2005}.
 In what follows we focus on large $N$,
and dropping $O\left(1/N\right)$ corrections we write $T=\frac{L}{N}$.
At very long times, the system will reach a distribution 
\begin{equation}
P({\mbox{dominant type at }}a)=\frac{e^{-E_{a}/T}}{\sum_{b}e^{-E_{b}/T}}\label{equil}
\end{equation}
Finding the stationary distribution has
been reduced to the solution of the \emph{equilibrium} Random Energy
Model \cite{derrida_random-energy_1980}, (see \cite{neher_emergence_2013}
and \cite{cammarota_spontaneous_2014}). In particular, we conclude
that, depending on the value of $N$, at very long times the system
will equilibrate to either a `liquid' phase (for $N/L<\ln2$) or a `frozen'
phase (for $N/L>\ln2$), see Fig. \ref{remsketch}: in the former random
extinction events stop the system from converging to the optimum level
of fitness, while in the latter this level is at long times reached.
{\em A feature we find here, and is a general fact, is that even
if the dynamics satisfy detailed balance, and are hence able in principle
to equilibrate, this takes place at unrealistically long times.}

The qualitative features of the population at different times has
long been known, the similarity of the role played by fluctuations
due to finite population size with thermal fluctuations has also been
noted long ago \cite{peliti_introduction_1997,crow_introduction_1970}.
Here the analogy becomes an identity, and 
the effect of accumulation of deleterious mutations  becomes just the question of an ordinary order-disorder phase
transition. Similarly, the effect of population bottlenecks becomes
the same as a spike in temperature.

\section{The phase diagram of the Darwinian SAT models}

The example of the House of Cards model suggest that we consider a
 phase diagram with variables $\propto 1/N$ and $ 1/\tau_0$. To obtain a meaningful phase diagram (Fig \ref{phasediagram}),
the scalings with growing $N$ must be consistently
defined. It is easy to see that for this one must keep  $N/L$ constant, and numerics further show that mutation times must
scale as $\tau \equiv \tau_0/ N$, where $\tau$ is constant for a given value of $N/L$ \footnote{This entails
that the width of the fitness distribution in the population at a
given time is $\sigma_{\lambda}^{2}\sim\frac{N}{L^{2}\tau_0}\sim\frac{1}{N^{2}}$
(following arguments as in \cite{kessler_evolution_1997,ridgway_evolution_1998} ). The  time-scale for a given spin flip is on average 
 $\tau_{point}=L\tau_0$. which  corresponds also to the time-scale for an individual to shuffle its entire configuration.}. Barrier-crossing mechanisms
 for the entire population are expected for both high  and flat and wide barriers \cite{weissman,Iwasa}.

\begin{figure}
\centering 
\subfloat[]{\includegraphics[angle=-90,width=0.32\columnwidth]{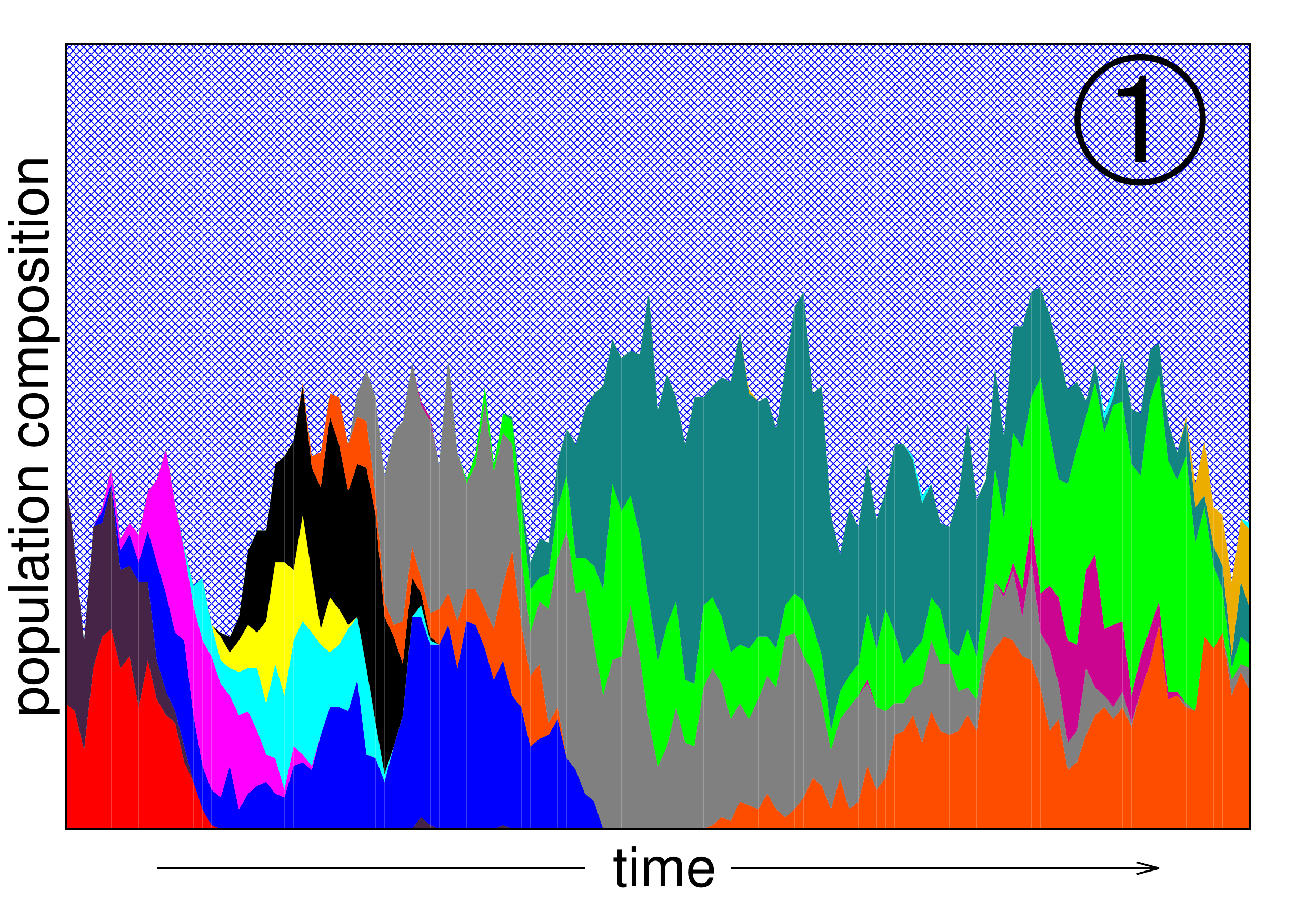}\label{concu1a}}
\subfloat[]{\includegraphics[angle=-90,width=0.32\columnwidth]{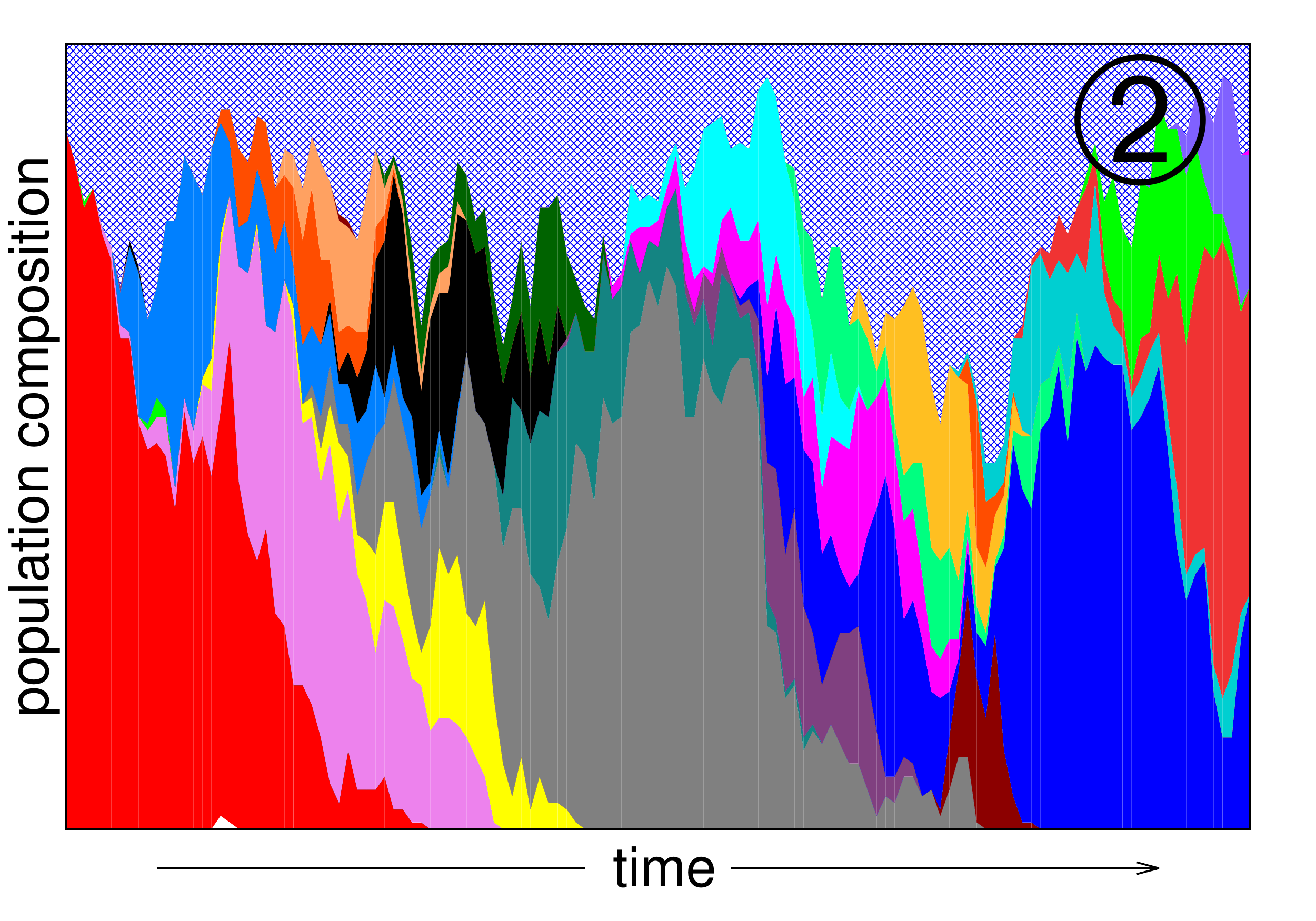}\label{concu1b}}
\subfloat[]{\includegraphics[angle=-90,width=0.32\columnwidth]{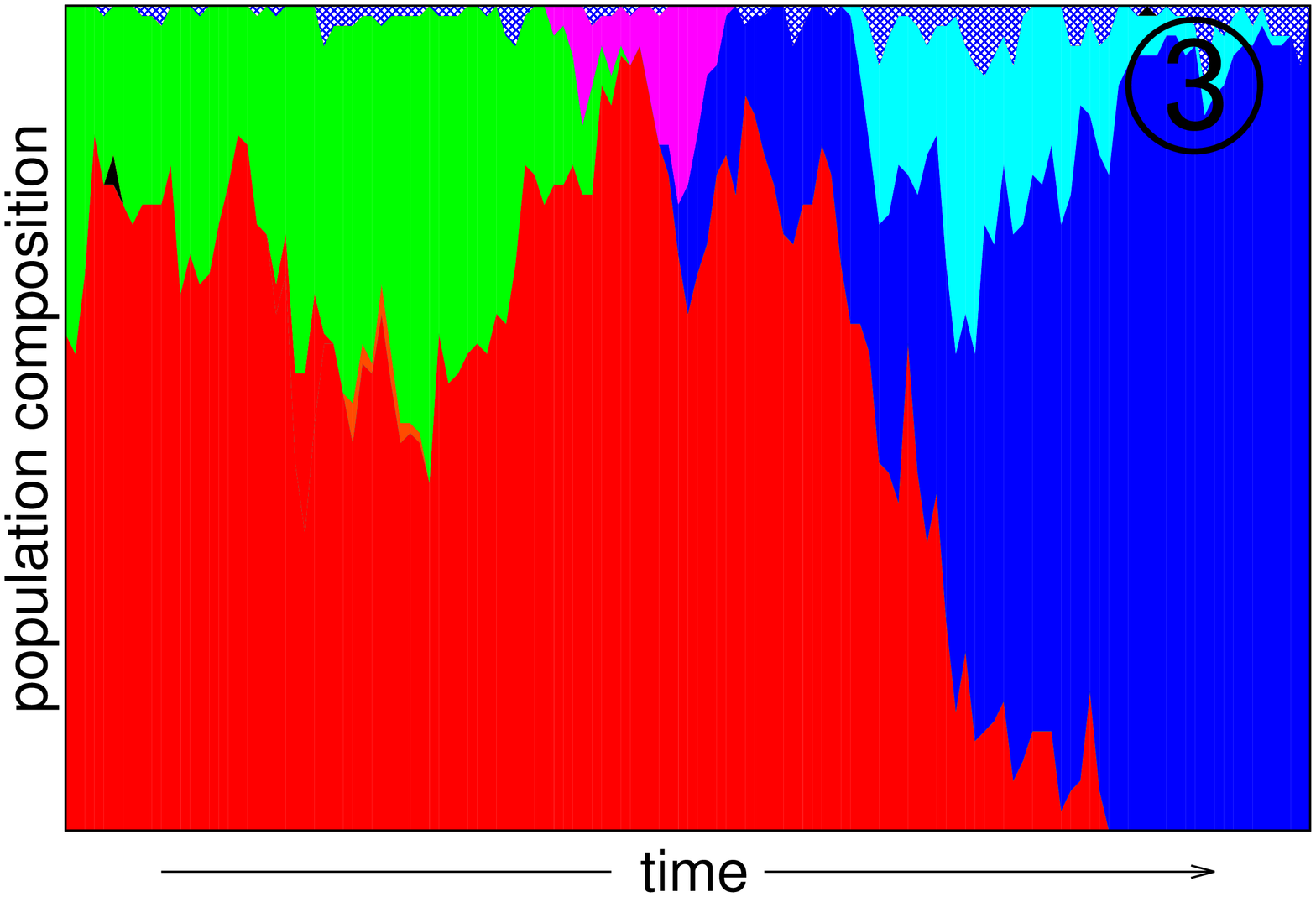}\label{concu1c}}
\protect\protect\protect\protect\caption{Mutant frequencies vs. time for the model in section \ref{XorSatExample}. Different
colors identify different mutants, and blue mesh includes all types
that never reach 10\% of the total population size. In (c), the population
at almost all times is dominated by a single mutant, whose identity
is replaced on rapid, successive sweeps. Detailed-balance is known
to hold here. In (a+b) more complex patterns are observed. Here too,
as we show below (section \ref{XorSatExample}), detailed-balance holds, provided that
proper averaging on short-times is applied. The labels of the figures
help locating them on the phase diagram of Fig \ref{phasediagram}.}
\label{concu1} 
\end{figure}

\subsection{ The thermal correspondence at low mutation rate}

\begin{figure}
\centering \includegraphics[angle=0,width=0.4\columnwidth]{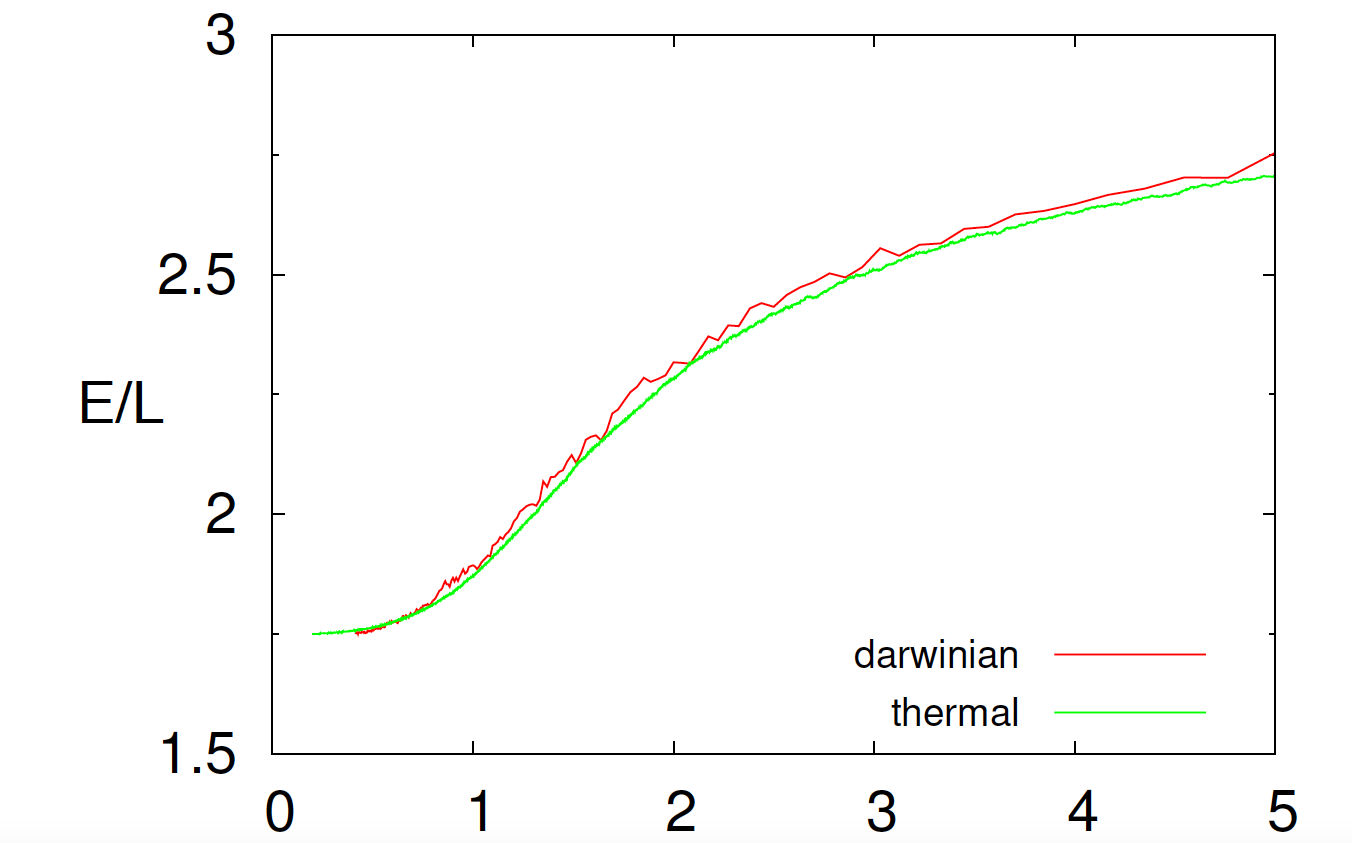}
\protect\protect\protect\protect\caption{Thermal versus Darwinian annealing for XorSAT, for $\tau=16$. Similar
results are obtained for K-SAT. Darwinian annealing is performed by
controlling the population so that it increases slowly, at the same
rate as in the corresponding thermal annealing with $T=L/N$.}
\label{annealing} 
\end{figure}

Let us first discuss the line $1/\tau=0$. In this regime, mutations are very rarely proposed, and  the system 
eventually falls  in
 the successional regime ($N/\left(\lambda\tau\right)\ll1$). 
As we have seen above, detailed balance is then expected to hold with temperature $T=\frac{L}{N}$. Indeed,
in Fig. \ref{annealing} we show the results of a simulated annealing
performed with an ordinary Monte Carlo program on a single sample,
superposed with a `populational annealing' performed by slowly increasing
the population of a set of individuals performing diffusion and reproducing
according to the fitness in Eq. (\ref{ftn}). The coincidence of both curves in terms of $T=L/N$ is reassuring.

We note in passing that the `thermal'
analysis allows one to make an evaluation of `genetic algorithms'
-- in this case we understand that the Darwinian Annealing will have
the same strengths and weaknesses as has Simulated Annealing. Furthermore,
we see that allowing for large populations from the outset may be
as catastrophic as is a sudden quench in an annealing procedure. 

In the limit $\tau\rightarrow \infty$, when the population behaves like a  thermal realization of a SAT system at temperature $T=\frac{L}{N}$,
the situation  is well understood for  the XorSAT and the  SAT problems \cite{krzakala_gibbs_2007}:
there is a (dynamic) glass transition at a  certain temperature
$T_{d}$ below which the phase-space breaks into components, and
the dynamics become slow, rendering the optimization  very hard.  This transition happens before the thermodynamic one, which itself 
is closely analogous to the freezing one of the REM.  

We may locate the dynamic 
transition by plotting the autocorrelation functions $C(t,t')=\frac{4}{L}\sum_{i}(s_{i}\left(t\right)-\frac{1}{2})(s_{i}\left(t'\right)-\frac{1}{2})$
at decreasing temperatures. As $T_{d}$ is approached from above,
the correlation decays in a two-time process, a fast relaxation to
a plateau followed by a much slower `$\alpha$-relaxation' (in the glass terminology),  taking a
time $t_{\alpha}$. As $T_{d}$ is reached, $t_{\alpha}$ diverges.
Below $T_{d}$, the system {\em ages}: the time $t_{\alpha}$ now
keeps increasing with time, $C(t,t')$ decays in a time $(t-t')_{decay}\sim t_{\alpha}(t')$
with $t_{\alpha}(t')$ an increasing function of $t'$.
What we have described is the `Random First Order Transition' \cite{kirkpatrick_scaling_1989}.
Nothing new here, as the system is equivalent to a thermal system,
known to exhibit such a transition. 

\begin{figure}
\includegraphics[angle=0,width=5cm]{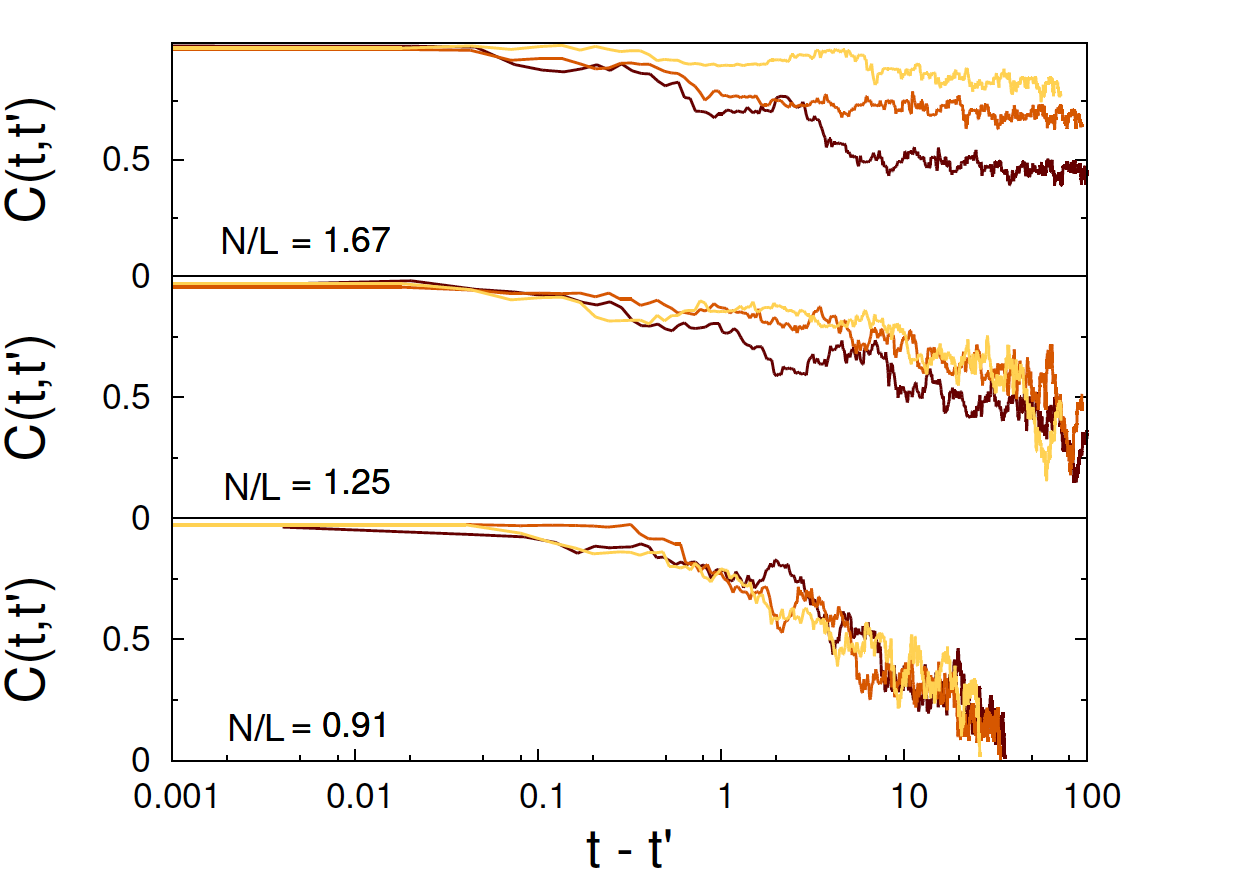}\includegraphics[angle=0,width=5cm]{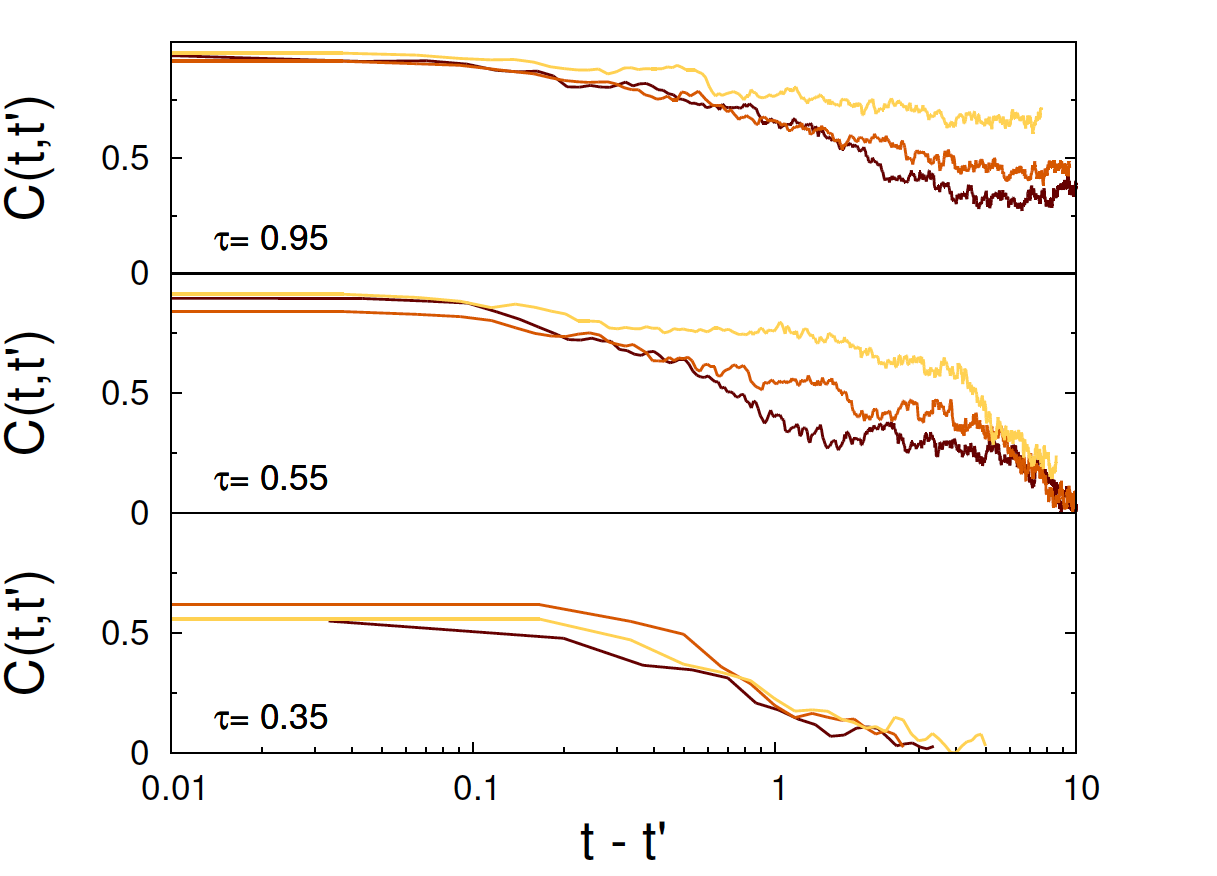}
\protect\protect\protect\protect\caption{Values of the $C(t,t')$ for a population crossing the glass transition
from liquid (bottom) to glass (top). Values are $\tau=5$, $N/L={0.91,1.25,1.67}$
(left), and $N/L=4$, $\tau={0.35,0.55,0.95}$. The transition line
is thus crossed by increasing $N$ or $\tau$, see the two arrows
in fig. \ref{phasediagram}, the points corresponding to the figures
are marked by crosses. The color code indicate the value of $\ln(t')$,
growing linearly from darker to lighter curves. In the top figures
$t_{\alpha}$ grows as the system ages. Here and in all the following
figures, the time is measured in units of mutation times $\tau$
(i.e. an individual  performs on average $O(L)$ flips in $\Delta t=O\left(1\right)$).
{The correlations are reasonably smooth, for a big
system, }{\emph{even for a single run}}}
\label{autocorr} 
\end{figure}

\begin{figure}
\centering \includegraphics[angle=0,width=0.5\columnwidth]{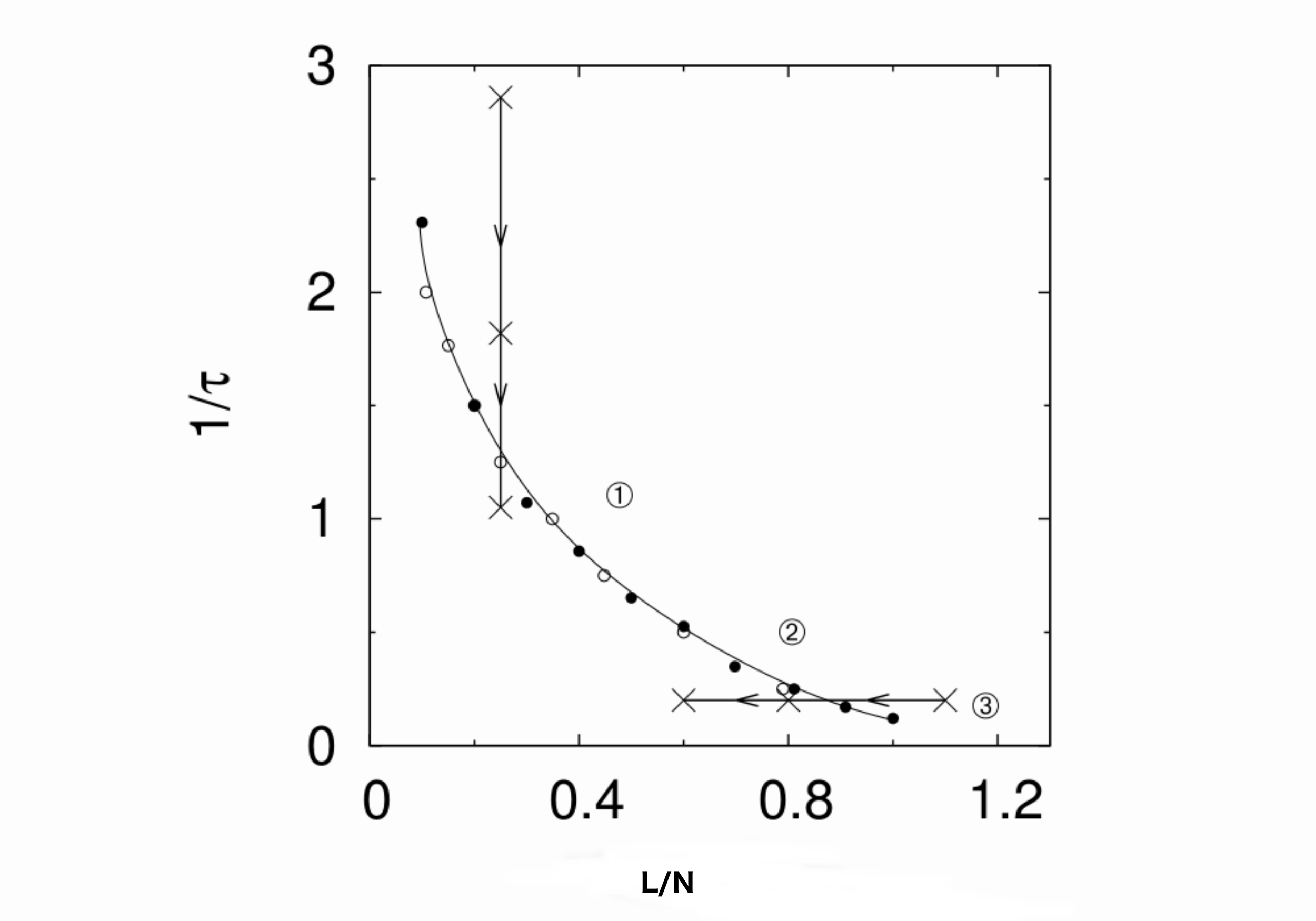}
\protect\protect\protect\protect\caption{Phase transition diagram in the $L/N$ vs $1/\tau$ plane. The transition
line is given by the black circles, which correspond to the points
where the value of the maximum in the $\chi_{4}$ function diverges,
approaching the line by changing $N$ (filled circles) or $\tau$
(empty circles). The two arrows correspond to two possible crossing
of the transition line, changing $\tau$ or $N$, the points indicated
by a cross being those represented in Fig \ref{autocorr}. Points
labeled by numbers correspond to the situations of Fig \ref{concu1},
and indicate the parameters used in the verification of the \emph{averaged}
detailed balance, Fig \ref{FT}.}
\label{phasediagram} 
\end{figure}

Let us now consider smaller $\tau$,
so that we no longer can assure that the $N$ individuals are fully
clustered in a configuration at most times, and $N$ no longer has an obvious thermal meaning. 
We approach the transition
by increasing $N$ at fixed $\tau$, and also by decreasing $\tau$ at
fixed $N$. The correlation curves obtained are shown in Fig (\ref{autocorr}):
the nature of the transition remains the same, but the transition
value of critical $N$ shifts with $\tau$. All in all, we obtain \cite{NS} the phase diagram of Fig \ref{phasediagram}.

\begin{figure}
\centering \includegraphics[angle=270,width=0.6\columnwidth]{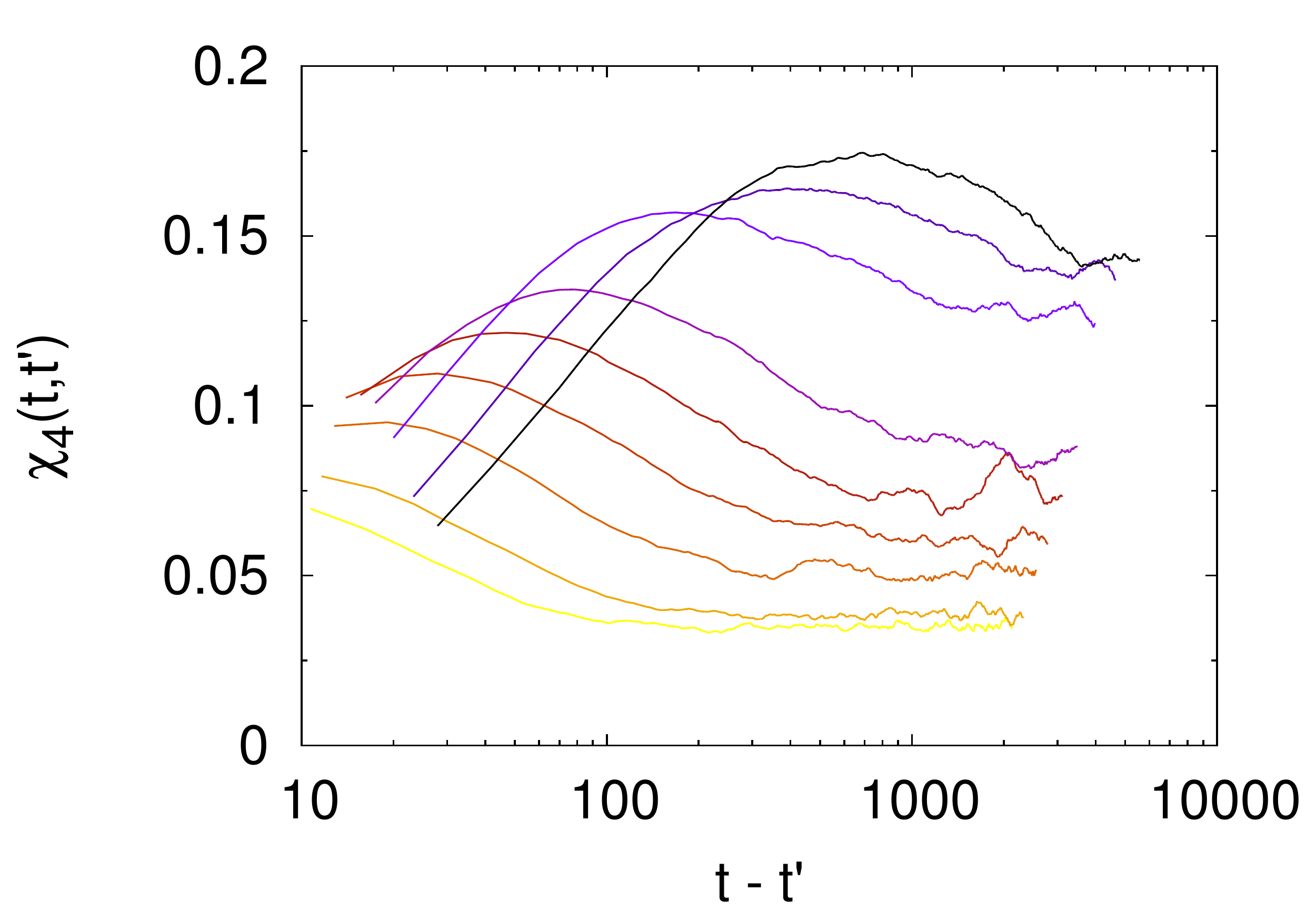}
\protect\protect\protect\protect\caption{$\chi_{4}$ versus time on approaching the transition. Data shown
for $N/L=1.4$, and a mutation rate $\tau^{-1}$ increasing linearly
from 0.25 to 0.65 (bottom to top, dark to light curves). At lower
mutations rates (not plotted) the glassy phase is reached, and the
curves do not present any maximum and keep increasing with the waiting
time $t$, as $t_{\alpha}$ is no longer defined.}
\label{chi4} 
\end{figure}

A confirmation of this is obtained by plotting the autocorrelation
`noise' (also known as dynamic heterogeneity)
\begin{equation}
\chi_{4}=L\left\langle \left[\frac{4}{L}\sum_{i}(s_{i}\left(t\right)-\frac{1}{2})(s_{i}\left(t'\right)-\frac{1}{2})\right]^{2}\right\rangle -LC(t,t')^{2}
\end{equation}
a quantity that peaks at a level expected to diverge at the transition,
at a time that we may estimate as $t_{\alpha}$, see Fig~\ref{chi4}.

\vspace{.2cm}

{\bf Thermal properties of slow dynamics in the region with slow dynamics (large $t_\alpha$).}

\vspace{.2cm}

We know that the $\tau^{-1}=0$ axis is just equivalent to the thermal problem,
with temperature $L/N$. What can we say about clustering for smaller
$\tau$? Figure \ref{concu1} shows the contributions of different
configurations (the top uniform color corresponds to contributions
smaller than 10\% each). We see that for all but the highest $\tau$,
the system is in the {\em concurrent mutation regime} \cite{desai_beneficial_2007},
and the thermal correspondence, applied naively, breaks down.
Considering several examples with timescale separation, we have conjectured \cite{unpublished}  that whenever the $\alpha$ relaxation
time is large (near and below the transition), the correspondence
with a thermal system may still hold, but taken for quantities that
are averaged over a time of several $t_\alpha$, and considering two situations at time-separations much
larger than $t_{\alpha}$.

Checking detailed balance numerically is extremely hard. We use here a Fluctuation Relation \cite{mustonen_fitness_2010,evans_fluctuation_2002}, as an indirect test. 
Because this theorem requires to start from equilibrium, we are only in a position to
do the test close to the glass transition, where the time-separation is large enough, but not within, because then equilibration
becomes problematic. 
  We thus place ourselves just {\em
above} the transition (so that a stationary distribution might be
reached) but not far from it (so that $t_{\alpha}$ is large), the
circled numbers in Figure \ref{phasediagram}. We start with a system
in equilibrium at time $t=0$, and we switch on a field $E\to E-hA$,
where $A$ is any observable, in our case we choose $A=\sum_{i}s_{i}$.
After an arbitrary time $t$ we measure again the value of $A$ and
check the equation (see \cite{unpublished}) 
\begin{equation}
\ln{P[A(t)-A(0)=\Delta]}-\ln{P[A(t)-A(0)=-\Delta]}=\beta h\Delta\label{ii}
\end{equation}

\begin{figure}
\centering \includegraphics[angle=270,width=0.48\columnwidth]{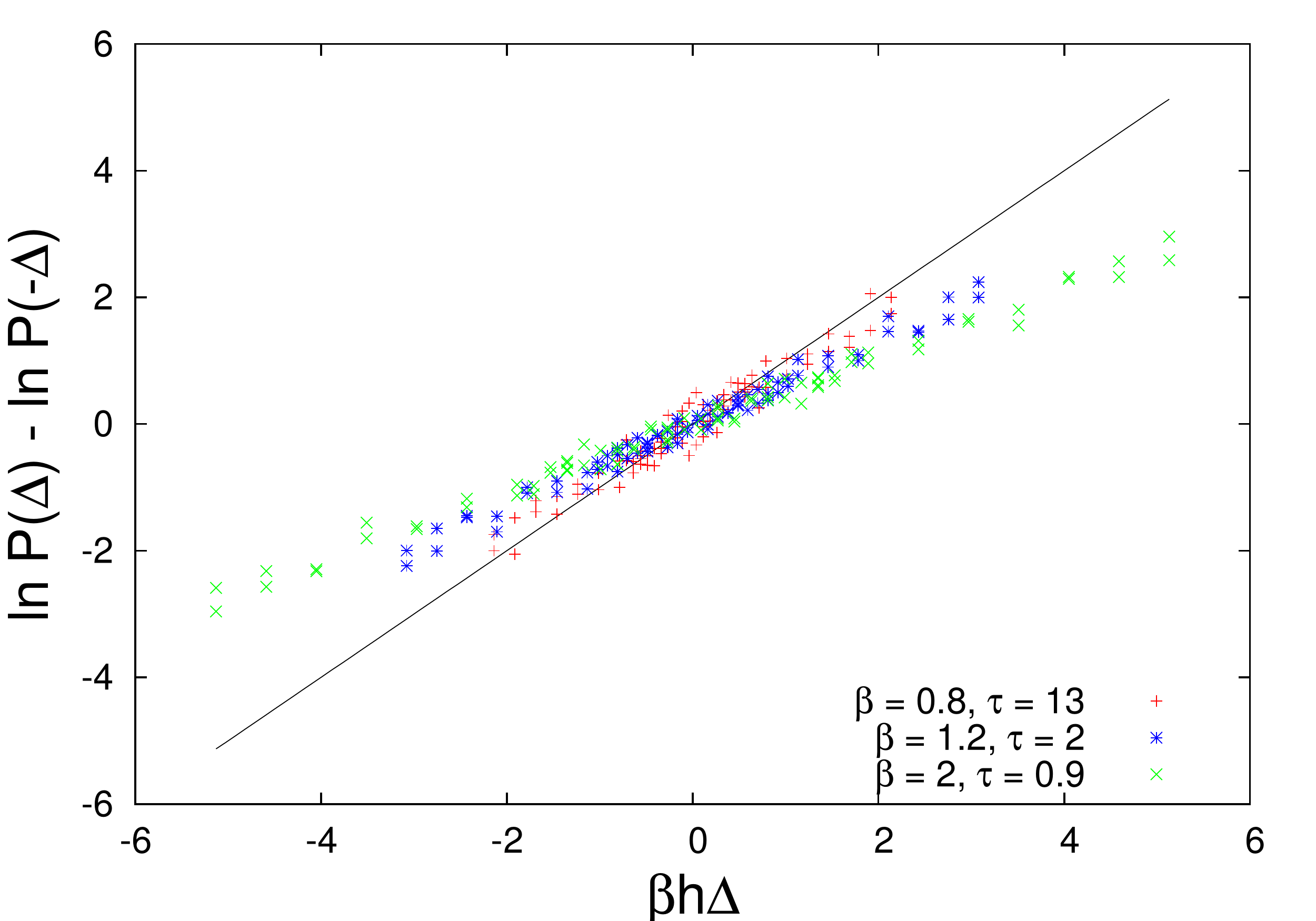}
\includegraphics[angle=270,width=0.48\columnwidth]{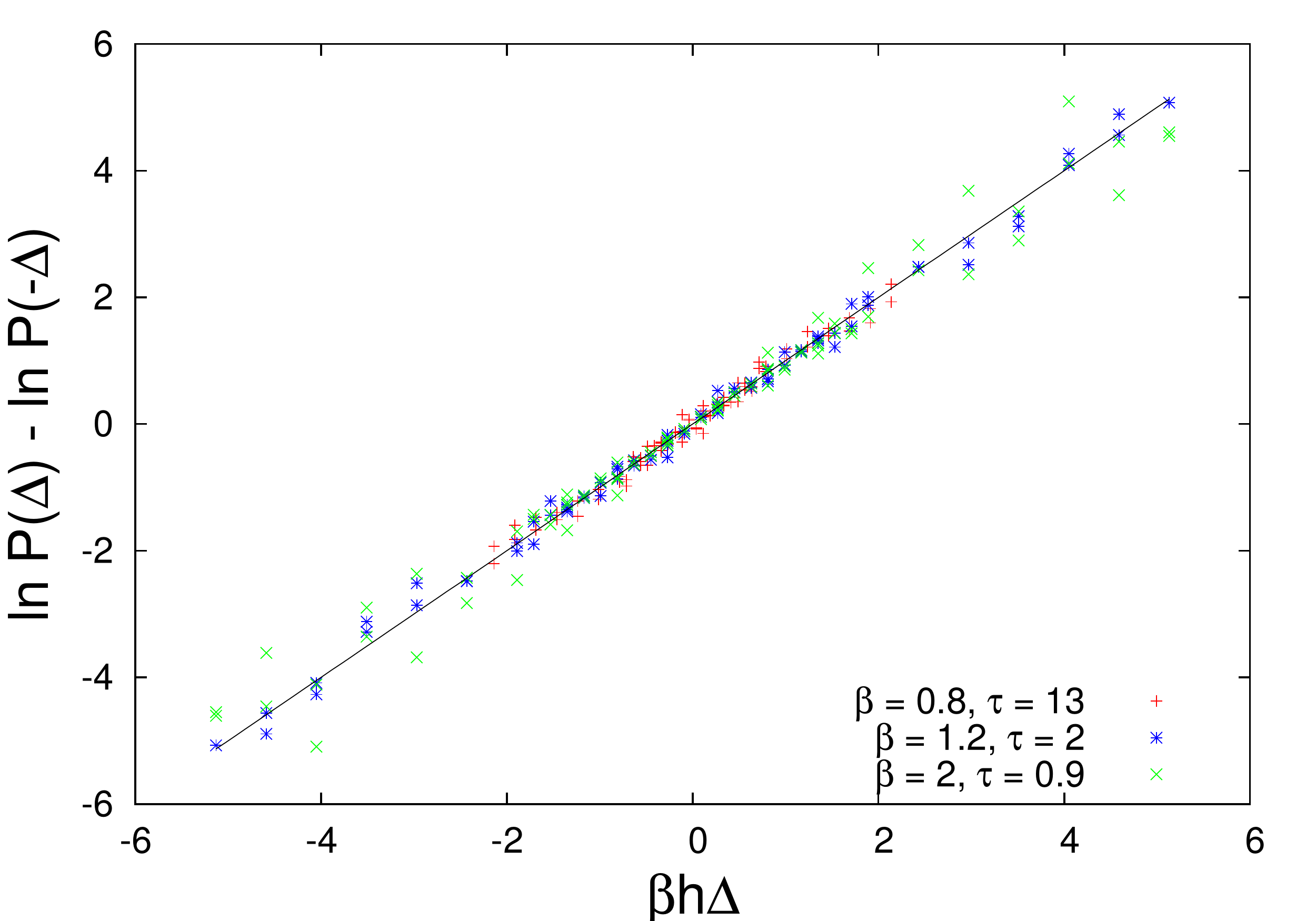}
\protect\protect\protect\protect\caption{$\ln P(\Delta)-\ln P(-\Delta)$ versus $\beta h\Delta$, for close times
and instantaneous values (left) and for quantities averaged over a
time-window, and times separated by more than $t_{\alpha}$ (right).
The latter works perfectly, without any fitting parameters. The parameters
for the three curves, $(\beta,\tau)={(0.8,13),(1.2,2),(2,0.9)}$, correspond
to those of figure \ref{concu1}, and are represented on the phase
diagram of figure \ref{phasediagram}.}
\label{FT} 
\end{figure}

We obtain the plot Fig \ref{FT} (left). It does not verify the
Fluctuation Theorem (\ref{ii}), thus showing that there
is no detailed balance or equilibrium, except for very large $\tau$.
This is what we expected, as there is no clustering into a single
type, and the connection with a thermal system {\em fails}. Instead, when
we compute the differences as $\Delta=\bar{A}(t)-\bar{A}(t=0)$, with
$\bar{A}$ the average of $A$ within a window comparable to the time
to reach a plateau -- the {\em `equilibration within a valley'} time $\ll t_{\alpha}$
-- and $t\sim3t_{\alpha}$, the relation (\ref{ii}) for the averaged
values $\Delta=\bar{A}(t)-\bar{A}(0)$ works perfectly, without fitting
parameters. 
{\em We interpret this as meaning that the jumps between valleys, taking a long time, are indeed
governed by a temperature $L/N$, although the diffusion inside a valley is not.}
%Finally, note that in our XorSAT simulations we have made two choices. First, we work
%with mutations that induce gradual jumps in the fitness (in sharp
%contrast, for example, with the House-of-cards model).  As a
%result, individual mutations incur a change in fitness $\delta\lambda_{mut}\sim1/N$.
%The jumps of $\delta\lambda_{mut}$ are larger than would allow for
%detailed balance to hold in a 1d step model, so a non-trivial extension
%of the known limiting conditions has been demonstrated. The scaling
%is the same as required by the step model, $\delta\lambda_{mut}\ll1/N$,
%and therefore here we have not shown that the mutation steps can be
%parametrically larger (as a function of $N$). The population width
%in our simulations is $\sigma_{\lambda}\sim1/N$, narrower than the
%minimum required for the step model: $\sigma_{\lambda}\sim N^{-1/2}$.

\subsection{An analogy with dense active matter}

\vspace{0.2cm}

Active matter is composed by particles that have a source of energy
other than a simple thermal bath. This may be the case of bacteria,
of specially defined propulsed particles, and of randomly shaken particles.
A way to model this situation, is to consider a source of energy in
the high frequencies (for example a thermal bath at high temperature
$T_{fast}$ with white noise) and a mechanism of dissipation, such
as a thermal bath with low-frequency correlated noise at a lower temperature
$T_{slow}$ (see Ref. \cite{berthier_non-equilibrium_2013}). The
high frequency source need not be a true thermal bath, it could consist
for example of random `kicks'. In any case, the system is not truly
in equilibrium, because for this one needs a same temperature at all
timescales. If one considers a situation with high density and a
suitable energy balance situation, the particle system approaches
a glass transition just as an ordinary thermal one. Timescale-separation
appears, the fast motion is essentially ruled by the `fast' excitation,
while the slow motion by the `slow' bath. One expects then detailed
balance at temperature $T_{slow}$ to hold for slow evolution and
time-averaged quantities, while it does not hold inside a state. The
situation is quite close to the one discussed in this paper, the role
of the `fast' excitation being played here by the fluctuations due
to the Darwinian nature of the dynamics within a state. The parameter
$1/\tau$ plays the role of the input energy rate at high frequencies,
while $\frac{L}{N}$ plays the role of $T_{slow}$. Our phase diagram
Fig. \ref{phasediagram} is strikingly similar the one of the active
matter model in Ref \cite{berthier_non-equilibrium_2013}.

\vspace{0.5cm}

\subsection{Changing environments: connection to glassy rheology}

\vspace{0.5cm}

A system as the one we are considering, which is achieving better fitness
by slowly adapting to a complex landscape, is extremely sensitive
to changes in this landscape. This effect has been discussed in \cite{mustonen_molecular_2008},
although in a slightly different form, and also in  \cite{kussell_phenotypic_2005}. The counterpart in glass physics
of this fact has long been known. Consider the situation \cite{struik_rejuvenation_1997}
of a plastic bar prepared a time $t_{w}$ ago from a melt. The polymers
constituting the bar slowly rearrange -- ever more slowly -- to energetically
better and better configurations, and this process is known to go
on at least for decades. The bar is out of equilibrium, a fact that
we may recognize by testing its response to stress, which measurably
depends on $t_{w}$. Now suppose that we apply a large, fixed deformation
to the bar, for example applying a strong torsion one way and the other. The new constraints change the problem of optimization
the polymers are `solving': we expect evolution to restart to a certain
extent, and the apparent `age' of the bar to become smaller than $t_{w}$.
This is indeed what happens \cite{struik_rejuvenation_1997}, a phenomenon
called `rejuvenation'. Rejuvenation brings about an acceleration in
the dynamics. If the changes are continuous and different,  instead of aging (growth of $t_{\alpha}$), the system
settles in a value of $t_{\alpha}$ that depends on -- adapts to --
the speed of change of the energy landscape. Note that this property of evolution speed, {\em as measured 
from the changes in the population} adapting
to landscape change speed, that is sometimes attributed to a form of criticality \cite{kauffman_metabolic_1969}, here  appears as a universal
property of aging systems.

\begin{figure}
\centering \includegraphics[width=6cm,angle=-90]{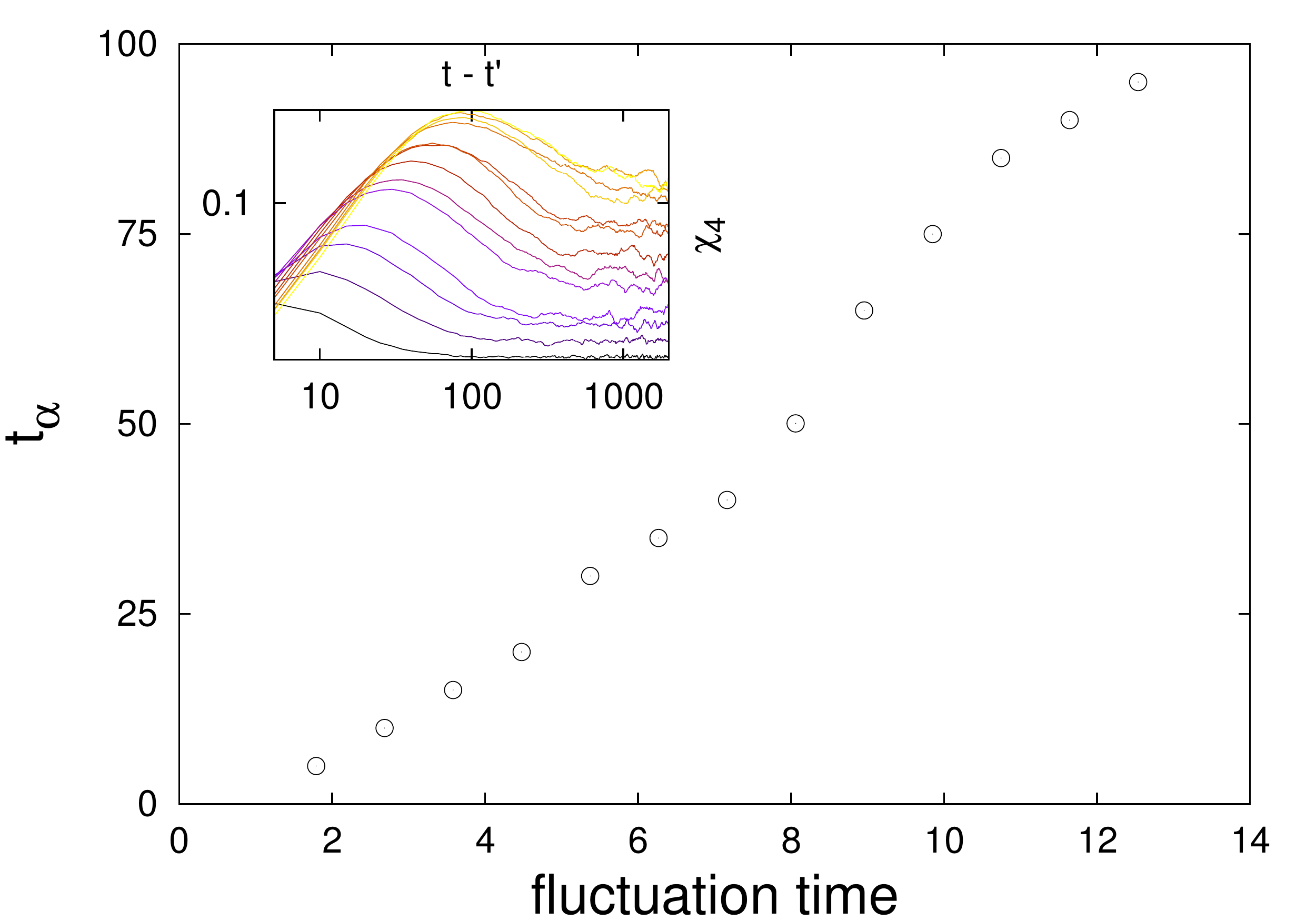} \protect\protect\protect\protect\caption{$\alpha$-time versus speed of variation of fitness landscape. Inset:
$\chi_{4}$ versus time for different speeds of random variation of landscape
(darker curves correspond to faster variations). The $t_{\alpha}$
value for each point of the main figure is the time at which the corresponding
curve reaches its maximum, a measure of the $\alpha$-time. }
\label{figa} 
\end{figure}

 Applying the same logic to our model, one expects a similar result. In order to model
 the changes in fitness landscape, we change at fixed intervals of time a randomly chosen clause, for example by changing the identity of one of the
 intervening Boolean variables (a slight change in the fitness function). For different rates of change, we plot the `age' of the system,
 as measured by $\chi_4$.  The results are shown on Figure
\ref{figa}: if the environment is randomly
changing, the system  evolves to accommodate various conditions,
and time-scales for changes in the environment are  reflected in
the time-scales for changes inside system.

%Although we shall not pursue this line here, let us remark that one need not consider only random changes of fitness landscape, but
%also repetitive ones. Recently, Fridman et al \cite{fridman_optimization_2014}
%subjected bacterial populations to intermittent exposure to antibiotics.
%All strains adapted via phenotypic changes and developed of tolerance
%by adjusting the lag time of bacteria before regrowth to match the
%duration of the antibiotic-exposure interval. In other words, the
%system adapted to the cycle itself. Correspondingly, in another recent
%paper, Fiocco\emph{ et. al.} \cite{fiocco_encoding_2014} have studied
%the effect of letting evolve a glassy system under the influence of
%a periodic field, strong enough to affect substantially its evolution.
%The system `adapts' to this non-stationary situation, just as it would
%to a stationary field: subjecting further the sample to new cycles
%of different amplitudes (`reading') one may easily distinguish the
%cycle at which it has been optimized (`writing'). 

\section{Conclusions }
\label{Conclusions}

We have studied the dynamics of a population of random walkers reproducing with a rate 
corresponding to a `rugged' fitness function. The natural phase variables are the size of the system
and the diffusion constant (the mutation rate). We obtain a glassy region in this phase diagram where the systems 
`ages', slowly evolving to ever better fitnesses. 
We have found evidence that even for relatively high diffusion constants, the size of the system may be interpreted
as an inverse temperature, provided one considers time-averaged quantities.
The phase diagram bears a striking resemblance to the one that would be obtained for the same system in
contact with a bath at  temperature $T$ 
driven simultaneously at the high frequencies, the intensity of the latter playing the role of the inverse mutation rate
$1/\tau$.

\vspace{.7cm}

\acknowledgments We would like to thank JP Bouchaud, and D.A. Kessler
for helpful discussions.

%\bibliographystyle{IEEEtran}
%\bibliography{IEEEabrv,paper}

\end{document}